\documentclass[twocolumn,showpacs,preprintnumbers,amsmath,amssymb]{revtex4}

\usepackage{graphicx}
\usepackage{dcolumn}
\usepackage{bm}

\begin{document}

\preprint{PRESAT-7803}

\title{Geometry and Conductance of Al Wires Suspended between Semi-Infinite Crystalline Electrodes}

\author{Tomoya Ono and Kikuji Hirose}
\affiliation{Graduate School of Engineering, Osaka University, Suita, Osaka 565-0871, Japan}

\date{\today}

\begin{abstract}
We present a first-principles study of a coherent relationship between the optimized geometry and conductance of a three-aluminum-atom wire during its elongation process. Our simulation employs the most definite model including semi-infinite crystalline electrodes using the overbridging boundary-matching method [Phys. Rev. B {\bf 67}, 195315 (2003)] extended to incorporate nonlocal pseudopotentials. The results that the conductance of the wire is $\sim$ 1 G$_0$ and the conductance trace as a function of electrode spacing shows a convex downward curve before breaking are in agreement with experimental data.
\end{abstract}

\maketitle
Atomic-scale wires, which can be produced by scanning tunneling microscopy and break-junction techniques, are the objects of considerable attention \cite{datta-ruitenbeek}, since the properties of such minute systems with low dimensionality can be different from those of bulks; as the size of the structure becomes smaller than the electronic mean free path, the electron transport becomes ballistic and the conductance is dominated by the electronic structure. To date, there have been a great deal of geometrical for revealing the mechanical and electrical behavior of such minute wires. Among the most exciting discoveries was that a single strand of gold atoms suspended between electrodes manifests conductance quantization in steps of G$_0$=2$e^2/h$ \cite{rubio-takayanagi-kizuka} as electrode spacing is increased, where $e$ is the electron charge and $h$ is Planck's constant. Furthermore, other studies showed that an atomic-scale wire made of aluminum atoms exhibits more complicated and even more interesting features; that is, the plateaus of conductance steps are not flat but have a positive slope before the wire breaks \cite{scheer-krans-rubio,sakai}. This somewhat counterintuitive result implies that the conductance increases even though the electrodes are pulled apart.

On the theoretical side, there are many first-principles studies for examining electronic structure and transport properties of atomic-scale wires using tight-binding approximation and/or the structureless jellium electrodes \cite{lang1,lang2,kobayashi1,kobayashi2,okamoto,tsukamoto,wan,brandbyge,palacios}, some of which reported results consistent with experimental data. For more strict theoretical predictions on details of electron conduction properties, it would be mandatory to completely eliminate artifacts and employ a realistic model where electrodes consist of {\it truly semi-infinite crystals}. Recently, the overbridging boundary-matching (OBM) formalism \cite{fujimotonano,fujimoto}, which is a novel first-principles treatment of electron transport properties for nanoscale junctions sandwiched between bona-fide semi-infinite crystalline electrodes, has been developed.

In this Letter, we implement elaborate first-principles calculations with the incorporation of the OBM formalism to elucidate a close relationship between the relaxed geometrical structure and electronic conductance of the 3-aluminum-atom wire during the elongation process using a model in which the wire is suspended between two semi-infinite Al(001) bulks. To the best of our knowledge, this is the first theoretical investigation of this relationship where the optimized aluminum wire is connected to genuine semi-infinite aluminum crystalline electrodes which are gradually separated. We find that this 3-aluminum-atom wire has a conductance of $\sim$ 1 G$_0$, and there is only one channel that gives a dominant contribution to the electron conduction at the Fermi level. The conductance trace as a function of electrode spacing decreases in the early stage of the elongation process and shows a minimum; in the case of shorter electrode spacing, a loop current (LC) is induced around the center atom of the bent wire so as to boost electron transmission. With an increase in electrode spacing, the LC weakens and the conductance reduces. When electrode spacing is still further increased, the wire forms into a straight structure with a one-dimensional character, which contributes to the enhancement of electron transmission. As a result, the curve of the conductance trace is convex downward before the wire breaks, and subsequently the conduction channels between the two electrodes disappear.

First, we determine the optimized atomic structure of the 3-aluminum-atom wire suspended between the Al(001) electrodes during its elongation. Our first-principles simulations are based on the real-space finite-difference method \cite{rsfd}, which enables us to determine both the self-consistent electronic ground state and the optimized geometry with a high degree of accuracy by virtue of the timesaving double-grid technique \cite{tsdg} and the direct minimization of the energy functional \cite{dmef}. The norm-conserving pseudopotentials \cite{ncps} of Troullier and Martins \cite{tmpp} are used to describe the electron-ion interaction. Exchange-correlation effects are treated by the local-density approximation \cite{lda} in the density-functional theory. The nine-point finite-difference case, i.e., the case of $N_f$=4 in Eq. (A2) (here and hereafter, the equation number refers to that in Ref. \cite{fujimoto}), is adopted for the derivative arising from the kinetic-energy operator in the Kohn-Sham equation. We set a cutoff energy of 25 Ry, which corresponds to a grid spacing of 0.63 a.u., and further take a higher cutoff energy of 224 Ry in the vicinity of the nuclei with the augmentation of double-grid points \cite{tsdg}.

For optimization, a conventional supercell, which is indicated by the rectangle enclosed with broken lines in Fig. \ref{fig:1}, is employed. The size of the supercell is chosen as $L_x$=$L_y$=15.12 a.u. and $L_z$=22.68 a.u. + 2$d$. Here, the $z$ axis is taken along the wire axis, the $x$ and $y$ axes are perpendicular to the wire, $L_x$, $L_y$ and $L_z$ are the lengths of the supercell in the $x$, $y$ and $z$ directions, respectively, and $d$ is the average value for the {\it projections} of interatomic distances between adjacent atoms of the wire onto the $z$ component. The wire is connected to the square bases with the side length of $\frac{a_0}{\sqrt{2}}$ at both its ends, modeled after the [001] aluminum strands, and all of these components intervene between the electrodes produced from three atomic layers of the Al(001) surface. The distance between the electrode surface and the basis as well as that between the basis and the edge atom of the wire is chosen to be 0.5$a_0$, where $a_0$ (=7.56 a.u.) is the lattice constant of the aluminum crystal. The position of the central atom of the wire is optimized on the (110) plane while the other atoms are fixed during the first-principles structural optimization. We find that the wire exhibits a bent structure below electrode spacing $L_{es}$ of 26.46 a.u. When elongated up to $L_{es}$=27.09 a.u., the wire manifests geometrical transition from a bent wire to a straight one, and finally distorts at $L_{es}$=28.35 a.u. \cite{comment1}

\begin{figure}[bt]
\vspace*{4.14cm}
\caption{Schematic representation of the model. The rectangle enclosed with broken lines represents the supercell used to evaluate the optimized atomic configuration and electronic structure.}
\label{fig:1}
\end{figure}

We next explore the electron transmission of the wire at zero-bias limit with the Landauer formula G=tr({\bf T}$^\dagger${\bf T})G$_0$, where {\bf T} is a transmission matrix. The global wavefunctions for infinitely extended states continuing from one electrode side to the other are evaluated by the OBM method \cite{fujimoto}. For the determination of the effective potential required in the OBM formalism, the electronic structure of the transition region which is composed of wire, bases, and a couple of atomic layers of electrodes (see Fig. \ref{fig:1}) as well as the electronic structure of the aluminum crystalline electrode is calculated self-consistently employing a standard supercell geometry with a three-dimensional periodic boundary. The wavefunctions in the crystalline electrode region are computed by the generalized eigenvalue equation of Eq. (19), and in order to circumvent a numerical instability caused by evanescent waves, the wavefunction-matching formula Eq. (11) at the interfaces between the electrode and transition regions is described in terms of the ratio matrices $R$'s, which are defined by Eqs. (25) and (B9). Furthermore, for calculating the conduction properties of the aluminum wires with $s$- and $p$-valence orbitals, we here extend the OBM scheme to the case including the {\it nonlocal} parts of the pseudopotentials of Troullier and Martins \cite{tmpp}; since the nonlocal parts are represented by block matrices in the Kohn-Sham Hamiltonian when the real-space finite-difference method is applied, the nine-point finite-difference formula for the kinetic-energy operator is adopted so that the off-diagonal elements of the finite-difference formula can entirely cover the block matrices of the nonlocal parts. Consequently, the Hamiltonian is described by a block-tridiagonal matrix composed of 4$N_{xy}$-dimensional block matrices, where $N_{xy}$ is the number of grid points on the $x$-$y$ grid plane \cite{comment2}.

Figure \ref{fig:2} shows the channel transmissions at the Fermi level as a function of $L_{es}$. The eigenchannels are investigated by diagonalizing the Hermitian matrix {\bf T}$^\dagger${\bf T} \cite{kobayashi2}. There are certain features that are common to the undistorted 3-aluminum-atom wire: throughout all the ranges of electrode spacing below $L_{es}$=28.35 a.u., the first channel, which has the characters of $s$ and $p_z$ for a straight wire, is widely open, while the second and third ones, which have the characters of $p_x$ and $p_y$ and are degenerate for the straight wire, have small transmission $\sim$ $1.0 \times 10^{-3}$, and the resultant electronic conductance is $\sim$ 1 G$_0$. Moreover, one can see that the conductance trace as a function of electrode spacing exhibits a convex downward curve having a minimum at $L_{es}$=26.46 a.u. Our result regarding the conductance trace accords with the available experimental data \cite{scheer-krans-rubio,sakai}, and in particular, is in excellent agreement with the data recently reported by Mizobata {\it et al.} \cite{sakai}.

\begin{figure}[bt]
\vspace*{4.66cm}
\caption{Channel transmissions at the Fermi level for the three-aluminum-atom wires as a function of electrode spacing.}
\label{fig:2}
\end{figure}

To gain an insight into this unique conductance trace, we plot in Fig. \ref{fig:3} the electron-density distributions and current densities of the first channel at the Fermi level for several electrode spacings. The states incident from the left electrode are depicted. In the case of a short electrode spacing such as $L_{es}$=24.57 a.u., the channel transmission is $\sim$ 1, and the first channel current forms an LC rotating around the center atom. Upon elongating electrode spacing, one finds that the LC weakens and consequently, the transmission exhibits a minimum. The electron transport through the wire is closely associated with the behavior of the LC: in fact, as the center atom is displaced upward from the most stable position depicted in Fig. \ref{fig:3}, the LC becomes enhanced to increase the transmission. When the wire is stretched up to electrode spacing of 27.09 a.u., where it forms into a straight wire from a bent wire, the transmission recovers $\sim$ 0.9. At this stage, the LC completely disappears and the current is axially symmetric around the wire axis. In the straight wire, the one-dimensional character strengthens, therefore the first channel transmission at the Fermi level is larger than that of the bent wire of $L_{es} \geq$ 25.83 a.u. After the wire's distortion, there are no conduction channels between the two electrodes.

\begin{figure*}[bt]
\vspace*{16.78cm}
\caption{(Color). Channel electron and current density distributions at the Fermi level for the three-aluminum-wire at (A) $L_{es}$=24.57 a.u., (B) $L_{es}$=25.83 a.u., (C) $L_{es}$=27.09 a.u., and (D) $L_{es}$=28.35 a.u. The upper panels show contours of the channel electron distributions, and the lower panels show arrows denoting the channel current distributions on the (110) plane. The states incident from the left electrode are shown. The large and small spheres indicate the atomic positions on and below the cross section, respectively.}
\label{fig:3}
\end{figure*}

One of the previous theoretical studies on the conductance of the 3-aluminum-atom wire employing structureless jellium electrodes observed a similar current density \cite{kobayashi2}. However, this simulation does not have sufficient accuracy to discuss strictly quantitative properties of electron transport because it was implemented using an inferior pseudopotential made only of local parts and a very restricted model where the bond lengths of the wire are fixed to be 5.4 a.u. For a definite interpretation of the relationship between geometry and conductance of the wire, a highly transferable pseudopotential having nonlocal parts and structural optimization are indispensable. Our result that the conductance is $\sim$ 1 G$_0$ is in accordance with that of the {\it ab initio} tight-binding calculation incorporating the square bases made of aluminum atoms \cite{palacios}. Compared with the other studies using nonlocal pseudopotentials and structureless jellium electrodes without any atomic layers or bases, most of the channel transmissions and conductance in the present study are smaller than those of the other studies: Lang \cite{lang1} and Kobayashi {\it et al.} \cite{kobayashi1} reported that the conductance of the wire is $\sim$ 1.5 G$_0$, and moreover, Kobayashi {\it et al.} reported the transmissions of the second and third channels are $\sim$ 0.3. In these calculations adopting jellium electrodes without any atomic layers, the channel transmissions and electronic conductance sensitively depend on direct effects from the jellium electrodes, and therefore the difference between our results and theirs may be mainly attributed to the absence of atomic layers or bases.

In summary, we have studied a close relationship between the optimized geometry and electronic conductance of the 3-aluminum-atom wire suspended between genuine semi-infinite aluminum crystalline electrodes. Our findings are that (1) the 3-aluminum-atom wire sandwiched between two Al(001) electrodes exhibits an electronic conductance of $\sim$ 1 G$_0$, (2) there is one conduction channel which widely opens at the Fermi level, and (3) the conductance trace as a function of electrode spacing behaves so as to draw a convex downward curve before the wire breaks. In the bent wire with a short electrode spacing of $L_{es} \leq$26.46 a.u., the electron transport property of the wire depends on the strength of the LC. On the other hand, in the case of a long electrode spacing of $L_{es} \geq$27.09 a.u., the one-dimensional character of the straight wire contributes towards the electron conduction to give rise to a conductance of $\sim$ 1 G$_0$. This result of the conductance trace is in accordance with experimental data. Work in progress is the study on the electron conduction of the aluminum wire attached to the electrodes having other crystal faces such as the (111) and (110) planes.

This research was partially supported by the Ministry of Education, Culture, Sports, Science and Technology, Grant-in-Aid for Young Scientists (B), 14750022, 2002. The numerical calculation was carried out by the computer facilities at the Institute for Solid State Physics at the University of Tokyo, and Okazaki National Institute.


\begin{references}
\bibitem{comment1}
We also carried out the structural optimization using a large supercell of $L_x$=$L_y$=21.40 a.u. and $L_z$=22.68 a.u. + 2$d$, and found no qualitative differences. This implies that the supercell lengths of 15.12 a.u. in the $x$ and $y$ directions taken in the text are adequate to ensure noninteraction of the wires in neighboring cells.
\bibitem{comment2}
In the present case of the aluminum wire, we choose the core radius of the nonlocal pseudopotentials to be 2.5 a.u., so that the core region contains 8 $x$-$y$ grid planes, and the matrix $B_z$ defined in Ref. \cite{fujimoto}, which corresponds to an offdiagonal block matrix in a block-tridiagonal form, becomes $4N_{xy}$ dimensional.
\end{references}
\end{document}